\documentclass[12pt]{iopart}

\usepackage{graphicx}
\usepackage{xcolor}
\usepackage{hyperref} 

\usepackage{multirow}

\begin{document}

\title[The human factor: results of a small-angle scattering data analysis Round Robin]{The human factor: results of a small-angle scattering data analysis Round Robin}

\author{Brian R. Pauw$^{1}$, Glen J. Smales$^{1}$}

\author{Andy S. Anker$^{2}$, Daniel M. Balazs$^{3}$, Frederick L. Beyer$^{4}$, Ralf Bienert$^{5}$, Wim G. Bouwman$^{6}$, Ingo Breßler$^{1}$, Joachim Breternitz$^{7}$, Erik S Brok$^{8}$, Gary Bryant$^{9}$, Andrew J. Clulow$^{10}$, Erin R. Crater$^{11}$, Frédéric De Geuser$^{12}$, Alessandra Del Giudice$^{13}$, Jérôme Deumer$^{14}$, Sabrina Disch$^{15}$, Shankar Dutt$^{16}$, Kilian Frank$^{17}$, Emiliano Fratini$^{18}$, Elliot P. Gilbert$^{19}$, Marc Benjamin Hahn$^{1}$, James Hallett$^{20}$, Max Hohenschutz$^{21}$, Martin Hollamby$^{22}$, Steven Huband$^{23}$, Jan Ilavsky$^{24}$, Johanna K. Jochum$^{25}$, Mikkel Juelsholt$^{26}$, Bradley W. Mansel$^{27}$, Paavo Penttilä$^{28}$, Rebecca K. Pittkowski$^{2}$, Giuseppe Portale$^{29}$, Lilo D. Pozzo$^{30}$, Paulo Ricardo de Abreu Furtado Garcia$^{31}$, Leonhard Rochels$^{15}$, Julian M. Rosalie$^{1}$, Patrick E. J. Saloga$^{32}$, Susanne Seibt$^{10}$, Andrew J. Smith$^{33}$, Gregory N. Smith$^{34}$, Venkatasamy Annadurai$^{35}$,
 Glenn A. Spiering$^{36}$, Tomasz M. Stawski$^{5}$, Olivier Tach\'e$^{37}$, Andreas F. Thünemann$^{1}$, Kristof Toth$^{38}$, Andrew E. Whitten$^{19}$, Joachim Wuttke$^{39}$}




\address{$^{1}$ BAM Federal Institute for Materials Research and Testing \\ 12205 Berlin, Germany}
\ead{{brian.pauw@bam.de}}

\address {$^{2}$ Department of Chemistry, University of Copenhagen, 2100 Copenhagen, Denmark} 
\address {$^{3}$ Institute of Science and Technology Austria (IST Austria), Am Campus 1, Klosterneuburg, 3400, Austria} 
\address {$^{4}$ U.S. Army Research Laboratory, 6300 Rodman Road, Aberdeen Proving Ground, Maryland 21005 (USA)} 
\address {$^{5}$ BAM Federal Institute for Materials Research and Testing, Richard-Willstätter-Str. 11, 12489 Berlin, Germany} 
\address {$^{6}$ Delft University of Technology} 
\address {$^{7}$ Helmholtz-Zentrum Berlin für Materialien und Energie GmbH}
\address {$^{8}$ FORCE Technology, Park Alle 345, 2605 Brøndby, Denmark} 
\address {$^{9}$ Physics, School of Science, RMIT University, Melbourne, Australia} 
\address {$^{10}$ ANSTO - Australian Synchrotron, 800 Blackburn Road, Clayton 3168, VIC, Australia} 
\address {$^{11}$ Virginia Polytechnic Institute and State University, Blacksburg, VA, USA 24060} 
\address {$^{12}$ Univ. Grenoble Alpes, CNRS, Grenoble INP, SIMAP, F-38000 Grenoble, France} 
\address {$^{13}$ Department of Chemistry, Sapienza Università di Roma, P.le A. Moro, 5, I-00185 Rome, Italy} 
\address {$^{14}$ Physikalisch-Technische Bundesanstalt, Abbestr. 2, 10587 Berlin} 
\address {$^{15}$ Department für Chemie, Universität zu Köln, Greinstraße 4-6, 50939 Köln, Germany} 
\address {$^{16}$ Department of Materials Physics, Australian National University, ACT, Australia} 
\address {$^{17}$ Faculty of Physics and CeNS, Ludwig-Maximilians-Universität München, Geschwister-Scholl-Platz 1, 80539 Munich, Germany} 
\address {$^{18}$ Department of Chemistry “Ugo Schiff” \& CSGI, University of Florence, Via della Lastruccia, 3, 50019 Sesto Fiorentino, Italy } 
\address {$^{19}$ Australian Centre for Neutron Scattering, Australian Nuclear Science and Technology Organisation, New Illawarra Road, Lucas Heights, NSW 2234, Australia} 

\address {$^{20}$ Department of Chemistry, School of Chemistry, Food and Pharmacy, University of Reading, Whiteknights, PO Box 224, Reading RG6 6AD, UK} 
\address {$^{21}$ Institute of Physical Chemistry, RWTH Aachen University, Landoltweg 2, D-52056 Aachen, Germany} 
\address {$^{22}$ School of Chemical and Physical Sciences, Keele University, UK} 
\address {$^{23}$ Department of Physics, University of Warwick, CV4 7AL, United Kingdom} 
\address {$^{24}$ APS, ANL, 9700S Cass ave, Lemont, IL, USA} 
\address {$^{25}$ Heinz Maier-Leibnitz Zentrum, Technische Universit{\"a}t M{\"u}nchen, D-85748 Garching, Germany} 
\address {$^{26}$ Department of Materials, University of Oxford, Parks Road, Oxford, OX1 3PH, UK} 
\address {$^{27}$ National Synchrotron Radiation Research Center, 101 Hsin-Ann Road, Hsinchu Science Park, Hsinchu 30076, Taiwan, R.O.C} 
\address {$^{28}$ Department of Bioproducts and Biosystems, Aalto University, P.O. Box 16300, FI-00076 Aalto, Finland} 
\address {$^{29}$ University of Groningen (NL)} 
\address {$^{30}$ University of Washington, Chemical Engineering, Box 351750, Seattle WA 98195-1750} 
\address {$^{31}$ Brazilian Synchrotron Light Laboratory (LNLS) -  Giuseppe Máximo Scolfaro 10000, zip: 13083-100, Campinas, São Paulo, Brazil} 

\address {$^{32}$ Independent Researcher} 
\address {$^{33}$ Diamond Light Source Ltd., Harwell Science and Innovation Campus, Didcot, Oxfordshire, OX11 0DE, United Kingdom} 
\address {$^{34}$ ISIS Neutron and Muon Source, Science and Technology Facilities Council, Rutherford Appleton Laboratory, Didcot, OX11 0QX, United Kingdom} 
\address {$^{35}$ University of Mysore, NIE First Grade College, Mysore - 570008} 
\address {$^{36}$ Department of Chemistry, Macromolecules Innovation Institute (MII), Virginia Tech, Blacksburg, VA 24061, USA} 
\address {$^{37}$ Universit\'e Paris-Saclay, CEA, CNRS, NIMBE, 91191, Gif-sur-Yvette, France} 
\address {$^{38}$ Materials Science and Engineering Division, National Institute of Standards and Technology, Gaithersburg, Maryland 20899, United States} 
\address {$^{39}$ Forschungszentrum Jülich GmbH, JCNS-MLZ, Lichtenbergstraße 1, 85747 Garching, Germany} 

\vspace{10pt}

\begin{abstract}

A Round Robin study has been carried out to estimate the impact of the human element in small-angle scattering data analysis. Four corrected datasets were provided to participants ready for analysis. All datasets were measured on samples containing spherical scatterers, with two datasets in dilute dispersions, and two from powders. Most of the 46 participants correctly identified the number of populations in the dilute dispersions, with half of the population mean entries within 1.5\,\% and half of the population width entries within 40\,\%, respectively. Due to the added complexity of the structure factor, much fewer people submitted answers on the powder datasets. For those that did, half of the entries for the means and widths were within 44\,\% and 86\,\% respectively. This Round Robin experiment highlights several causes for the discrepancies, for which solutions are proposed. 

\end{abstract}

%
\vspace{2pc}
\noindent{\it Keywords}: Round Robin, small-angle scattering, nanostructure quantification
%
\submitto{\MSMSE}
%
\maketitle
%
%

\section{Introduction}

The scientific method has been historically developed to eliminate human and instrumental bias from understanding of the natural world. It has been applied to a wide range of fields with various levels of success. This success may be gauged using tools such as Round Robin (RR) experiments, where, for example, identical objects are circulated to various laboratories, enabling the quantification of the spread in findings. Ideally, the observations and resulting conclusions are independent of the observer or instrumentation, provided the means and methodology meet a minimum standard. If this is the case, we can be confident that the results are free of bias, and likely to be an accurate representation of the object or phenomenon under investigation. 

Several notable RR experiments in nanomaterial analysis compare the results of different techniques (insofar that different techniques are able to provide truly comparable end parameters). To more closely capture the true human or instrumental variability, however, experiments focusing on a single technique or even a single aspect of a technique are perhaps better suited. Such focus allows us to pinpoint the larger contributors to inter-laboratory variability, with the eventual goal of eliminating or minimizing such dependencies. Notable examples of such studies have been carried out in fields such as atom probe tomography \cite{Dong-2019}, X-ray diffraction \cite{Madsen-2001, Scarlett-2002}, neutron scattering \cite{Rennie-2013}, neutron powder diffraction \cite{Whitfield-2016}, Bio-SAXS (Biology-specific Small-angle X-ray Scattering) \cite{Trewhella-2022}, and surface area determination following the Brunauer, Emmett and Teller method \cite{Osterrieth-2022}.

Along this vein, a large RR experiment was carried out several years ago focusing on the collection of small-angle scattering data for nanoparticle liquid suspensions. Small-angle scattering is a technique for (traceable) quantification of nanostructures in bulk amounts of sample. After appropriate data correction, information might be retrieved on the scatterer morphology (``form factor''), its size distribution \protect{\footnote{While the data contains information proportional to the mass or volume of the scatterers, for narrow distributions this can be converted to number-weighted distributions while maintaining low uncertainties.}}), or its packing (``structure factor''). In most cases, one of these three may be elucidated upon the provision of information or assumptions on the other two. In rare cases, two or more of these pieces of information can be convincingly extracted from the data. 

In the aforementioned RR, the small-angle x-ray scattering datasets from a wide range of laboratories were collected and subsequently analyzed using a single set of programs with consistent starting parameters \cite{Pauw-2017a}. From this experiment, it was clear that at least for non-challenging samples, most laboratories and instruments were able to collect consistent data resulting in standard deviations on the order of a percent for the mean particle size, and about ten percent for the population size distribution width. The next logical step then, is to find out what influence the human factor would have on the analysis of the data. 

The influence of researchers on the results can be investigated by circulating a dataset to be interpreted, and quantifying the variation on the resulting morphological parameters \cite{Osterrieth-2022, Madsen-2001, Scarlett-2002}. In particular in small-angle scattering, the data analysis can be a stumbling block, and so the expectation is to see a large spread in the results for this Data Analysis Round Robin (DARR) for small-angle scattering. 

Given the wide range of possible samples and analyses in this field, the challenge was to find representative datasets that would be: 
\begin{enumerate}
  \item{} covering a range of relevant materials and common challenges for datasets
  \item{} of high quality to minimize result variation through data uncertainties
  \item{} from well-characterized and well-understood samples
  \item{} accompanied by the same nominal level of supplementary information as is normally provided by materials researchers
\end{enumerate}
Further complicating the experiment was a practical limitation on the answer-space: a machine-readable answer sheet needed to be developed that would present the resulting morphological parameters for variation analysis while limiting the telegraphing of a desired result or answer space. Several aspects of the answer sheet were deliberately left vague, to attempt to eke out further information on what the small-angle scatterer might understand for common but confusing terms (one example of this is the ``mean size'' of the spherical scatterers, not specifying whether diameter or radius was meant). Further discussion on the appropriateness of the answer sheet is provided below. 

While invariably constrained by the aforementioned considerations, the results inferred from the 46 entries nonetheless provide a good insight on the challenges facing small-angle scattering as a materials science tool. After presenting the datasets, methods and results, a brief discussion highlights the possible areas in need of further attention and avenues of improvement for the field, interpreted from the results. Additionally, a paragraph is spent on discussing possible improvements on future RR experiments. Lastly, it should be mentioned that while the results presented herein are necessarily limited, the anonymized results are available on Zenodo \cite{DARRdataset-2023}, as well as the Jupyter notebook used for the correction, interpretation and visualisation, in the hopes that alternative or extended interpretations of the results may be developed.

\section{Dataset Descriptions}

\begin{figure}[ht]
\begin{center}
\includegraphics[width=\textwidth]{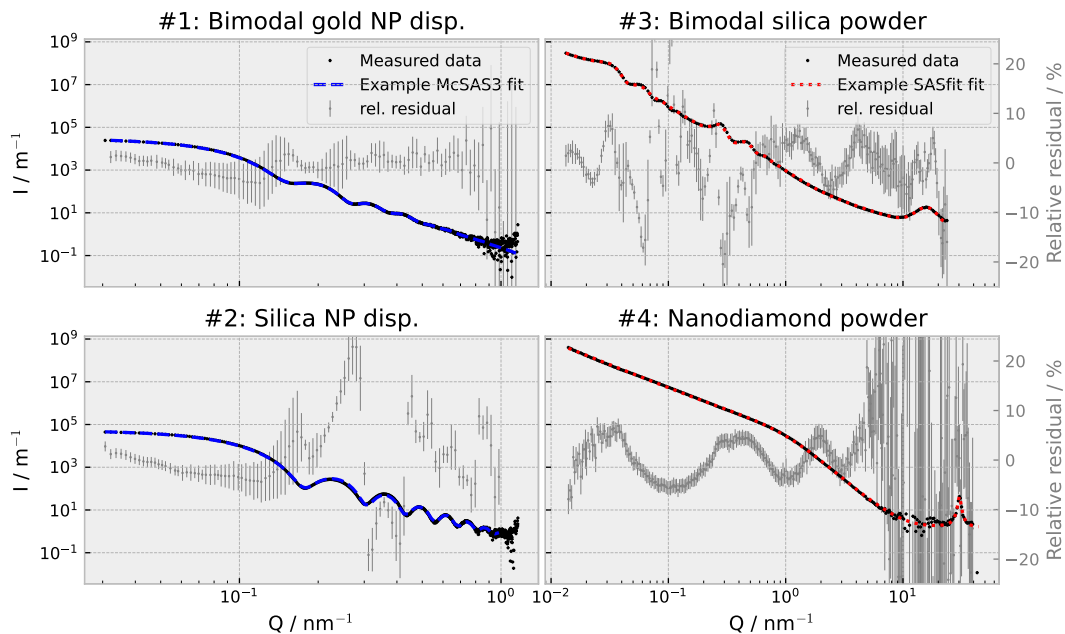}
\caption{The four datasets (black) chosen for this RR experiment, with example fits and relative residuals (light grey, relating to the secondary, right-hand side axis). The example fits have been generated using McSAS3 (left, blue) and SASfit (right, red). The size distributions resulting from these fits are shown in Figure \ref{fg:d1example}.}
\label{fg:datasets}
\end{center}
\end{figure}

Four datasets of one-dimensional scattering data, representative of two dilute and two dense nanoparticle systems, were made available to willing participants in the form of tabulated three-column .dat files, containing q, I(q), $\sigma_I(q)$. Where q is the scattering vector in units of $nm^{-1}$, I(q) is the scattering intensity in units of $(m \cdot sr)^{-1}$ and $\sigma_I(q)$ the absolute uncertainty of the intensity (one standard deviation). In addition, participants were provided with a letter describing the datasets, and the task ahead, as well as an Excel sheet for tracking results in a standardized form (see Zenodo repository \cite{DARR-2023}).

The four datasets are shown in figure \ref{fg:datasets}, with model fits that serve only as a suggestion of an appropriate model for the datasets. Datasets 1 and 2 were fitted in McSAS3 utilizing a spherical form factor, whilst datasets 3 and 4 were fitted in SASfit using spherical form factors (with log-normal distributions) with appropriate structure factors (sticky hard sphere and mass fractals for datasets 3 and 4 respectively), alongside background contributions and peak functions to better describe the wide-angle data. These fits can also be found in the Zenodo repository \cite{DARRdataset-2023}, a select number of fitting parameters (mean and widths for each population of each dataset) are shown in Table \ref{tb:fitvals}.

Dataset 1 and 2 originate from publicly available measurements \cite{Deumer-2022} performed at the SAXS beamline of the Physikalisch-Technische Bundesanstalt (PTB) at the Berliner Elektronenspeicherring-Gesellschaft f\"ur Synchrotronstrahlung II mit beschr\"ankter Haltung (BESSY II m.b.H.) \cite{Krumrey-2001, Wernecke-2014}, on reference samples synthesized for the European Metrology Programme for Innovation and Research (EMPIR) ``nPSize'' project. These reference samples are detailed in \cite{Bartczak-2022}. 

Datasets 3 and 4 were measured as powders using the MOUSE (Methodology Optimization for Ultrafine Structure Exploration) \cite{Smales-2021}. X-rays were generated using a microfocus X-ray tube with a copper anode, followed by multilayer optics to parallelize and monochromatize the X-ray beam to an approximate wavelength of Cu K$\alpha$ ($\lambda$ = 0.154\,nm). Samples were mounted as small amounts of powder in a thin laser-cut holder and held in place between two pieces of low-scattering Scotch Magic Tape. Scattered radiation was detected on an in-vacuum Eiger 1M detector (Dectris, Switzerland), which was placed at multiple distances between 55 to 2507\,mm from the sample. The resulting data has been processed and scaled to absolute intensity using the DAWN (Data Analysis WorkbeNch) software package in a standardized complete 2D correction pipeline with uncertainty propagation \cite{Smales-2021, Pauw-2017}.

\subsection{Dataset 1: Bimodal gold nanoparticles}

Dataset 1 is from a material designated as nPSize 1. This sample was designed to contain two populations with known concentrations of spherical gold nanoparticles (NPs) in water, with diameters of (30 and 60)\,nm at a 1:1 number ratio. Given this number ratio, the volume fraction ratio is approximately 1:8. In other words, the smaller population (population 1) is present at a 1/8th volume fraction and therefore only contributes much less to the scattering pattern than the larger population (population 2). It should be noted that in practical measurements, the modality of the populations is often not known. The existence of the second population has therefore not been explicitly revealed to the participants but merely hinted at through the design of the answer form.

\begin{figure}[ht]
\begin{center}
\includegraphics[width=\textwidth, 
clip]{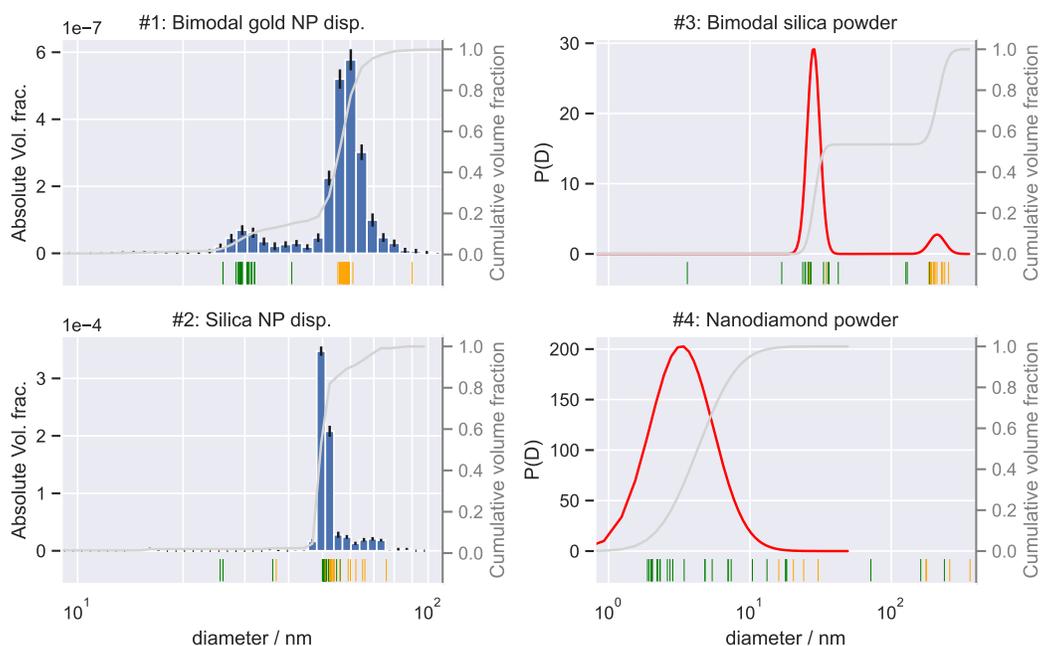}
\caption{Example volume-weighted distributions for datasets 1 through 4, using McSAS3 (left) and SASfit (right). The McSAS histogram bars show the integrated volume fraction for the span of that bar, the SASfit results show the volume-weighted probability function. For both, the cumulative distribution function (CDF) is plotted in grey on the secondary axis. The marks below indicate the participant-submitted population means for population 1 (green) and population 2 (orange).}
\label{fg:d1example}
\end{center}
\end{figure}

One possible solution for this scattering pattern is provided in Figure \ref{fg:d1example} using McSAS3 \cite{McSAS3}. This example solution shows the presence of two populations. When these populations are analyzed in the diameter range of $20 \leq D(\mathrm{nm}) \leq 40$ and $40 \leq D(\mathrm{nm}) \leq 80$, respectively, they provide the volume-weighted population means and widths (standard deviations) for populations 1 and 2 as shown in Table \ref{tb:fitvals}. For comparison, values from example fits using SASfit \cite{Kohlbrecher-2022} have also been added. Note that we cannot claim these solutions to be ``correct'', but they serve as an example. 


\begin{table*}[ht]
\caption{Means and widths determined from example fits to the data using McSAS and SASfit for datasets 1 to 4, and populations 1 and 2. Note the subscript $v$ for volume-weighted, and $n$ for number-weighted values. Uncertainties in brackets where available. \label{tb:fitvals}}
\small
\centering
\begin{tabular}{|c|c|c|c|c|c|c|} \hline
\textbf{Dataset} & \textbf{Population} & \textbf{Software} & \multicolumn{4}{c|}{\textbf{Results (nm)}} \\
\hline
 &  & & $\mu_v$ & $\mathrm{SD}_v$ & $\mu_n$ & $\mathrm{SD}_n$ \\
 \hline
 \hline
\multirow{4}{*}{1} & \multirow{2}{*}{P1} & McSAS & 30.8(4) & 3.8(4) & - & -  \\
 & & SASfit & 31.2 & 5.0 & 29.1 & 4.7  \\
 & \multirow{2}{*}{P2} & McSAS & 59.0(2) & 6.2(4) & - & -  \\
 & & SASfit & 58.8 & 5.3 & 57.4 & 5.2  \\
 \hline
 \multirow{3}{*}{2} & \multirow{2}{*}{P1} & McSAS & 52.2(6) & 4.4(1) & - & -  \\
 & & SASfit & 50.7 & 1.1 & 50.6 & 1.1  \\
 & P2 & SASfit & 56.1 & 8.5 & 52.4 & 7.9  \\
 \hline
 \multirow{2}{*}{3} & P1 & SASfit & 28.5 & 3.0 & 27.6 & 2.9  \\
 & P2 & SASfit & 214.8 & 27.4 & 204.5 & 26.1  \\
 
 \hline
 4 & P1 & SASfit & 4.26 & 2.20 & 1.92 & 0.99  \\

\hline
\end{tabular}
\end{table*}



\subsection{Dataset 2: Silica nanoparticles}

The second dataset is nPSize 10 from the same series, where the scatterers consist of a narrow distribution of nominally monomodal silica with a nominal diameter of 60\,nm \cite{Bartczak-2022}. Recent discussions revealed that this sample may also contain a minor fraction of a slightly larger population (c.f. Table \ref{tb:fitvals}). McSAS3 example fits (as well as SASfit, not shown here) do indicate the presence of a small fraction of a broad distribution of particles, and a significant number of participants found the same. 


\subsection{Dataset 3: Mixture of AS-40 and 250\,nm Silica powders}

Dataset 3 was produced in-house by mixing together two spherical silica materials in a 1:1 mass ratio. Smaller silica spheres were obtained by freeze-drying Ludox AS-40 (Sigma-Aldrich, ca. 22\,nm in diameter), whilst the larger spheres were synthesised using the St\"ober process, where tetraethyl orthosilicate (TEOS, Sigma-Aldrich, 98\,\%) was added to a solution of ethanol (Sigma-Aldrich, 96\,\%), water and ammonium hydroxide solution (ACS reagent, ca. 28\,\%) and left to stir at room temperature for 24 hours. The resulting suspension was then centrifuged and washed with ethanol before being dried at 60\,$^{\circ}$C overnight. SASfit example distribution is show in figure \ref{fg:d1example}, with means and widths detailed in Table \ref{tb:fitvals}.



\subsection{Dataset 4: Nanodiamond powder}

Dataset 4 was measured from a commercial sample of nanodiamonds obtained from PlasmaChem GmbH in Berlin, catalogue number PL-D-G02. These are globular diamond particles with a nominal diameter between (4 to 6)\,nm supplied as a dry powder. Some technical details and additional references demonstrating their use are available for this material on the PlasmaChem website \cite{PlasmaChem}. SASfit example distribution is show in figure \ref{fg:d1example}, with means and widths in Table \ref{tb:fitvals}.



\subsection{Electron micrographs}

\begin{figure}[ht]
\begin{center}
\includegraphics[width=\textwidth]{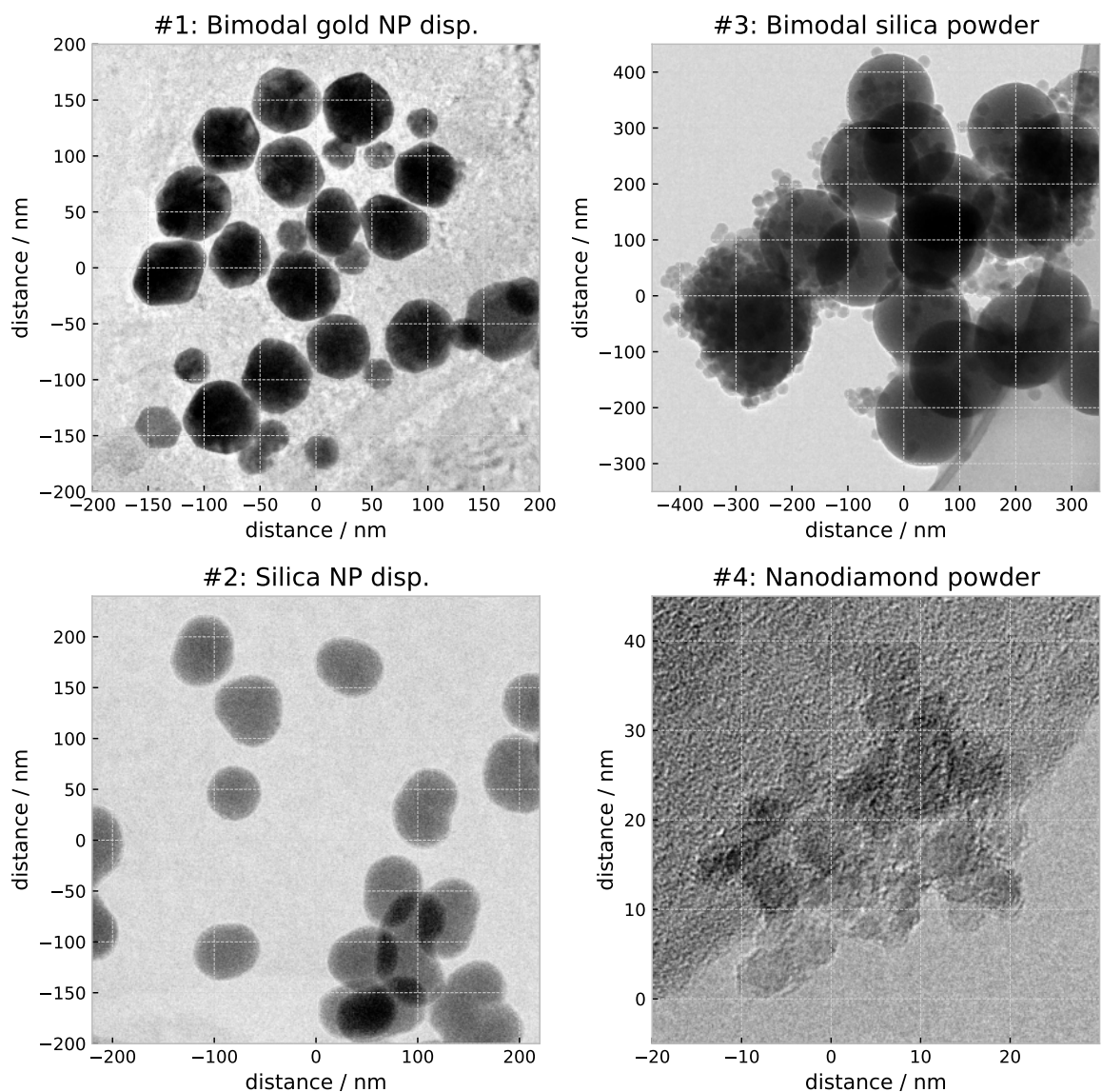}
\caption{Electron micrographs of the scatterers that make up the four datasets. The images from dataset 1 and 2 originate from the nPSize project repository \cite{Pollen-2021}. Images for the samples of dataset 3 and 4 were recorded on site.}
\label{fg:TEMs}
\end{center}
\end{figure}

Electron micrographs showing the scatterers underpinning the four datasets are shown in Figure \ref{fg:TEMs}. The images from dataset 1 and 2 were measured at the Commissariat \`a l'\'energie atomique et aux \'energies alternatives (CEA) and deposited in a Zenodo repository \cite{Pollen-2021}. 

Images for the samples of dataset 3 and 4 were recorded using an electron microscope available on site. For these two, the experimental details are as follows:
TEM samples were prepared by dispersing the powders via ultrasonication for a minimum of 5\,minutes in ethanol.
One to three droplets of the resulting suspensions were placed on Cu TEM grids coated with lacey- or holey-carbon films. TEM observations were conducted on a JEOL 2200FS instrument, operating at 200\,kV. Images were acquired in bright-field TEM mode, using high-contrast apertures to enhance the visibility of the particles, except for the approximately 10\,nm nanodiamonds of dataset 4, where high-resolution TEM images were obtained instead.


\section{Data read-in and corrections}

The 46 valid submitted answer sheets had to undergo several processing steps before they could be compared. Author information and reported analysis results were read in separately to aid anonymization. The following corrections were applied to the evaluation data in this order:
\begin{enumerate}
    \item Manual corrections were done to some sheets, to ensure the entries were in the right column for reading, to remove extraneous information, change decimal commas to periods, to fill in missing information (after communication with the author), etc.. This manual correction step was more frequently necessary than it perhaps should have been. 
    \item The software package names were sorted and shortened to the minimal identifying name. For software packages that were only used once or twice, they were categorized under ``Other'', in order to not compromise anonymization. 
    \item The weighting category was forced into either ``volume'', ``number'', or ``not defined''
    \item From the value range of the first population mean, it was determined whether the author had (most likely):
    \begin{enumerate}
        \item (72\,\%): interpreted ``size'' as ``radius'' (set correction factor to 2),
        \item (4\,\%): missed the dataset unit information (additional correction factor of 10)
        \item (4\,\%): mis-corrected the dataset units or reported information in \r{A}ngstr\"om (additional correction factor of 0.1)
        \item (40\,\%): Used the SasView software package, but reported the size distribution width in SasView's ``polydispersity'' units rather than in a population width in standard deviation (set correction factor to the mean radius). Other software packages might also report the distribution width in other ways, but this is not known and thus not corrected for. 
    \end{enumerate}
    \item Due to the limited number of entries, no outlier test is applied to further exclude submitted values. 
    \item Lastly, the entries of the set of concatenated data are randomized, so that they were no longer in the sequence in which they were ingested. 
\end{enumerate}

\section{Results and discussion}

\subsection{Overall statistics}

\begin{figure}[ht]
\begin{center}
\includegraphics[width=.59\textwidth]{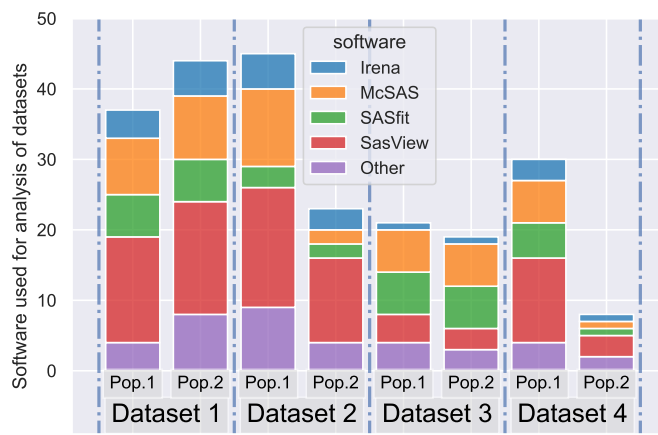}
\caption{The number of entries for each population of each dataset, which can be read off from the total height of each bar, subdivided in software. Datasets 1 and 3 were from samples with a nominally bimodal distribution.}
\label{fg:software}
\end{center}
\end{figure}

Not all of the datasets were fitted by all participants, and not all participants identified the same number of populations in the datasets. Unavoidably, it was telegraphed through the answer sheet that there might be more than one population present, however it was not indicated for which sample(s) that would be the case. Figure \ref{fg:software} shows how many participants had entered values for a given population for each dataset, and what software they used to detect this population. 

Most participants recognized a bimodal population in datasets 1 and 3, about half added a second population to dataset 2, and almost none saw a second population in dataset 4. Datasets 3 and 4 were more of a challenge than 1 and 2, with only about half of the participants entering results for these. 

The software packages used for these analyses consisted of four main packages, in alphabetical order\protect{\footnote{When software names are abbreviations, they have not been spelled out here for reasons of legibility and as they are known by their names, not their spelled-out (b)ac(k)ronyms.}}: ``Irena'' \cite{Ilavsky-2009}, ``McSAS'' \cite{Bressler-2015} and/or ``McSAS3`` \cite{McSAS3}, ``SasFit'' \cite{Kohlbrecher-2022} and ``SasView'' \cite{SasView-2022}. note that some packages such as Irena offer multiple methods for optimization. Other users used more uncommon software, among which (in alphabetical order): 
\begin{itemize}
    \item autosaxs (in-house software),
    \item GNOM via BioXTAS RAW \cite{Hopkins-2017}, 
    \item PySAXS \cite{pySAXS},
    \item SAXS-numerical inversion \cite{Bender-2017}. 
    \item XSACT \cite{XSACT},
\end{itemize}
as well as several in-house developed codebases. 

Given the range of challenges posed by the datasets, the analysis likewise could benefit from leveraging different approaches implemented in the software packages. Thus, Figure \ref{fg:software} shows the software packages used per dataset. A fairly even distribution between software packages is evident from this. This indicates a healthy ecosystem on the one hand, but complicates comparison as each software package may report parameters such as distribution widths, volume fractions, and goodness-of-fit in their own unique way. 

\subsection{Just a moment: on number- vs. volume-weighting}

It should be noted that the various software packages typically present the key population statistics based either on a number-weighted or on a volume-weighted distribution. A mean size, for example, can thus be expressed as the mean size by number, or the mean size by volume (or, more precisely, by mass). 

We can express these mathematically by using the definition of weighted sample moments as a basis. This allows us to define total amount (zeroth raw moment), mean (first raw moment), variance (second central moment), and any higher (central) moments $k>2$ (with increasing uncertainty) using the equations in Table \ref{tb:moments}:

\begin{table*}[ht]
\caption{Defining number- and volume-weighted moments $m$ of populations consisting of $N$ contributions. The subscripts $n$ and $v$ denote number- and volume-weighting, respectively. $D_i$ is a diameter of an object (sample) $i$, $v_{f, i}$ the volume fraction of the same, \label{tb:moments}}
\small
\begin{tabular}{| p{0.07\textwidth}| p{0.16\textwidth}| p{0.3\textwidth}| p{0.35\textwidth} |}
\hline
\textbf{k} & \textbf{meaning} & \textbf{number-weighted} & \textbf{volume-weighted} \\
\hline
\hline
$k=0$ & integral value & $m_{0,\mathrm{n}} = 1 $ & $m_{0,\mathrm{v}} = v_f = \sum_{i=1}^{N} v_{f, i}$ \\
\hline
$k=1$ & sample mean & $\mu_{\mathrm{n}} = \frac{1}{N}\sum_{i=1}^{N} D_i$ & $\mu_{\mathrm{v}} = \frac{1}{v_f}\sum_{i=1}^{N} D_i v_{f, i}$ \\
\hline
$k\geq 2$ & variance, skew, kurtosis, etc. & $m_{k,\mathrm{n}} = \frac{1}{N}\sum_{i=1}^{N} \left(D_i - \mu_\mathrm{n}\right)^k$ & $m_{k,\mathrm{v}} = \frac{1}{v_f}\sum_{i=1}^{N} \left(D_i - \mu_\mathrm{v}\right)^k v_{f, i}$ \\
\hline
\end{tabular}
\end{table*}

From these, the width $\sigma$ is obtained from the variance $m_2$ through: 
\begin{equation}\label{eq:variance}
\sigma = \sqrt{m_2}
\end{equation}

and an optional adjustment for sampling bias to obtain unbiased moments can be obtained through:
\begin{equation}\label{eq:biascorrection}
m_{k, \mathrm{unbiased}} = m_k \frac{N}{N-1}
\end{equation}

Maths aside, the practical difference between the two weightings is that a volume-weighted mean is always larger than the number-weighted mean, with increasing discrepancy for broader distributions (the other moments are also non-interchangeable as they define different population distributions). While previous studies found that the information in a small-angle scattering dataset closely represents a volume-weighted distribution \cite{Pauw-2013a}, there is nothing stopping analytical fitting methods from modeling a size distribution using number-weighted parameters, albeit with increasing uncertainty on the smaller end as the distribution broadens. As this uncertainty on the distribution is not normally shown in analytical modeling packages, a user can be led to believe that such a number-weighted distribution is determined with equal precision over the entire range. Results from this Round Robin, for example from the dataset 1 results (\textit{vide infra}), show that either a misconception exists on what the values presented by some software packages represent, or that it is unclear that volume-weighted and number-weighted parameters are inherently different.

\subsection{Dataset 1 as an example}

\emph{Note that as the entries contain a wealth of information, only a subset can be shown and discussed here. The reader is encouraged to download the anonymized results and accompanying Jupyter notebook themselves for further investigations \cite{DARRdataset-2023}.}

\begin{figure}[ht]
\begin{center}
\includegraphics[width=.6\textwidth]{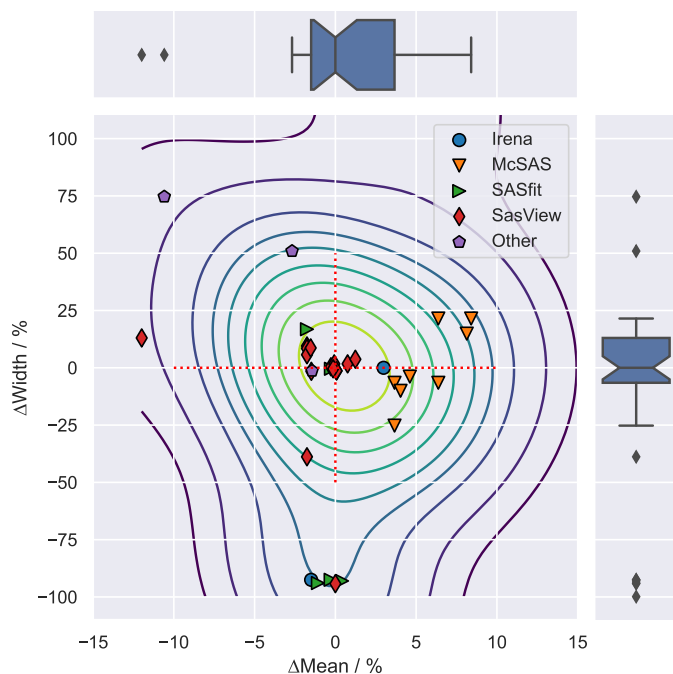}
\caption{The deviation from the median in percent, for both the mean and width of the smallest population in dataset 1, separated by software. The 2D kernel density estimate (KDE) is shown as contour lines, the median is highlighted with a red-dashed line cross, and the notched box plots show the quartiles for their respective mean or width values, with the whiskers indicating the 95\,\% confidence interval.}
\label{fg:bivariateKDE}
\end{center}
\end{figure}

Dataset 1 is, perhaps, the most straightforward, and thus serves well as a starting point for discussion of the actual results. Figure \ref{fg:bivariateKDE} attempts to show as much relevant information as possible in a single figure. As the size distribution is reasonably narrow, the volume-weighted mean and the number-weighted mean are sufficiently proximate to be shown on the same plot. 


The visualisation shows a breadth of entries that spans about $\pm$ 7\,\% for a 95\,\% confidence interval on the reported mean scatterer dimension, and a remarkable 50\,\% of the entries fall within 1.5\% of the median mean. The reported widths are deviating much more, with 50\,\% of the entries within 44\,\% of the median width. This is very likely due to the inconsistent reporting of distribution widths by the various software packages. Despite the instructions specifying that the widths should be reported as a standard deviation, many values were well outside the viable range for this specification, often hovering between 0 to 1. One intermediate conclusion from this is that, due to this reporting inconsistency, the reported widths therefore are largely unusable for the purposes of comparison. In lieu of an acceptable solution, we will thus concentrate mainly on the reported population means for the remainder of the paper. 

A second interesting aspect is the clustering of the various software packages. For example there is a cluster of McSAS results, slightly to the right of the mean. The cluster is offset slightly to the right, likely because of the volume-weighting of the results having an effect on the population means. That argument does not hold universally, however, with reported number- and volume-weighted values spanning the field.

Lastly, it is clear that a knowledge gap exists with the users of \emph{some} software packages vis-a-vis the weighting used for the reported population values (i.e., means and widths). This is evidenced by 44\,\% of SasView users indicating that the values are volume-weighted, against 51\,\% reporting that these represent number-weighted values (and a few hedging their bets and not reporting weighting at all). For SasFit, this is 52\,\% and 43\,\%, respectively. For the record, SasView and SASfit both report number-weighted population statistics, but SASfit plots the distribution in volume-weighted form per default, adding to the confusion. This appears to be a user interface (UI) issue, as no such confusion appears to exist with the users of McSAS and Irena, where 97\,\% and 78\,\% respectively reported the values as representing volume-weighted values.

\subsection{Findings on all datasets}

\begin{figure}[ht]
\begin{center}
\includegraphics[width=.99\textwidth]{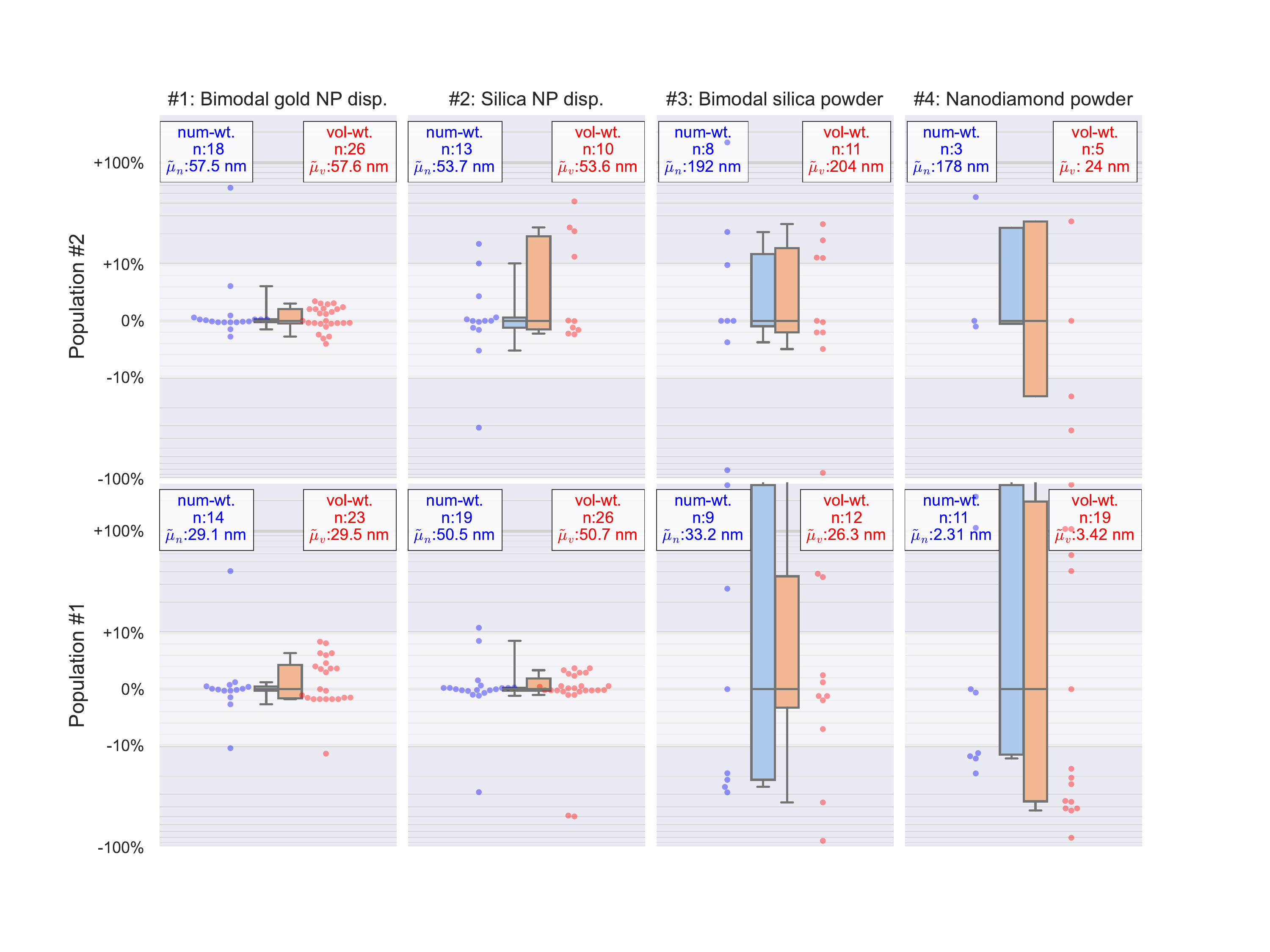}
\caption{The precision of the means submitted to the DARR. This is represented as a deviation of the reported means from the median mean in percent for each population for each dataset, separated by number- (blue) and volume- weighting (red), each referenced to their respective median means $\bar\mu_n$ and $\bar\mu_v$. This data is represented as a boxplot (showing quartiles, with the whiskers indicating the 95\,\% confidence interval) accompanied by the individual datapoints. The plots are shown on a symmetrical logarithmic scale with the lighter shaded region representing a linear region from $-10 \leq \bar\mu\% \leq 10$. Both populations of each dataset contains their absolute median mean and number of entries in the inset box. A larger spread of the submitted values can be observed for the more challenging powder-based samples compared to the dispersed samples.}
\label{fg:reldevmeans}
\end{center}
\end{figure}

When the relative spread of the submitted population means are compared for each population in each dataset, it becomes apparent that the more challenging powder-based samples exhibit a much larger spread (Figure \ref{fg:reldevmeans}). This can be attributed to the complications posed by the presence of a significant structure factor in dataset 3, and the near-fractal broadness of the distribution in combination with a structure factor underlying dataset 4. Changes in the chosen structure factor, or the structure factor (local) volume fraction parameter can significantly affect the determined means. Likewise, differences in size distribution models can equally impact the end result. 

\begin{figure}[ht]
\begin{center}
\begin{minipage}{0.48\textwidth}
    \begin{center}
    \includegraphics[width=\textwidth]{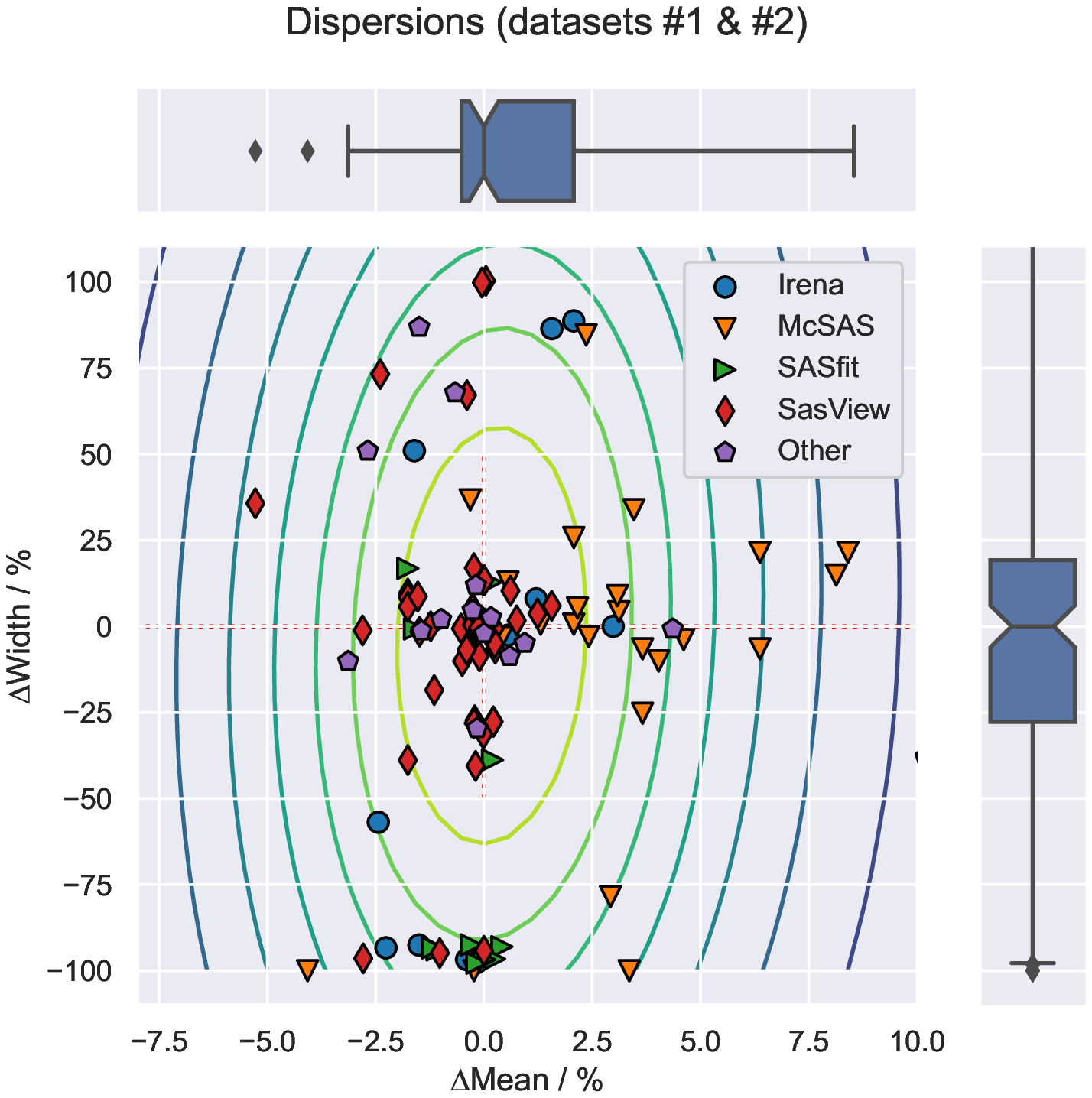}
    \end{center}
\end{minipage}\hfill
\begin{minipage}{0.48\textwidth}
    \begin{center}
    \includegraphics[width=\textwidth]{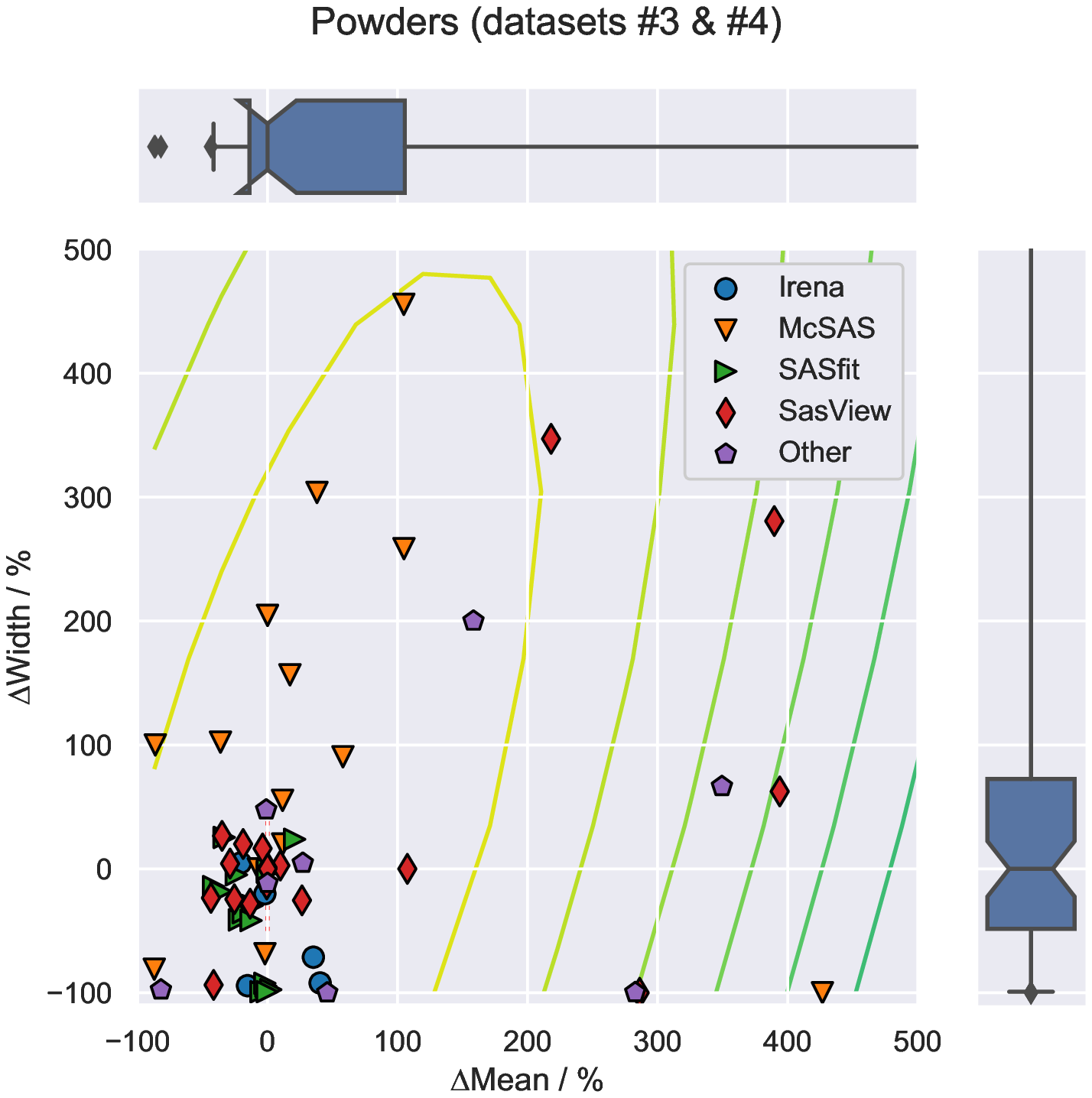}
    \end{center}
\end{minipage}
\caption{All entries for both populations of dataset 1 and 2 (left) and dataset 3 and 4 (right), shown as deviation from the median population mean (horizontal) and deviation from the median population width (vertical). Datapoints are grouped by software. The 2D KDE is shown as contour lines, the median is highlighted with a red-dashed line cross, and the notched box plots show the quartiles for their respective mean or width values, with the whiskers indicating the 95\,\% confidence interval. }
\label{fg:bivariateKDEseparated}
\end{center}
\end{figure}

This implies that a distinction could be made of the results between the deviations of entries of dataset 1 and 2, and of 3 and 4, respectively. Once this is done (Figure \ref{fg:bivariateKDEseparated}), we can conclude that for low-concentration dispersions, the 95\,\% confidence interval of the determination of the population means can be determined within about ten percent. The widths, however, are not as consistent between the participants, likely due to the aforementioned inconsistencies in the reporting between the various packages, and as it stands can vary by more than 100\,\%. The same analysis of the results for the powder samples shows an unusably broad distribution of results, indicating that blind intercomparisons between powder results of distinct laboratories may not yet be reliable. This could, perhaps, be improved by agreeing on a consistent approach for analysis of such samples (c.f. Paragraph \ref{sc:suggestions}).

\subsection{Volume fractions}

\begin{figure}[ht]
\begin{center}
\includegraphics[width=.99\textwidth]{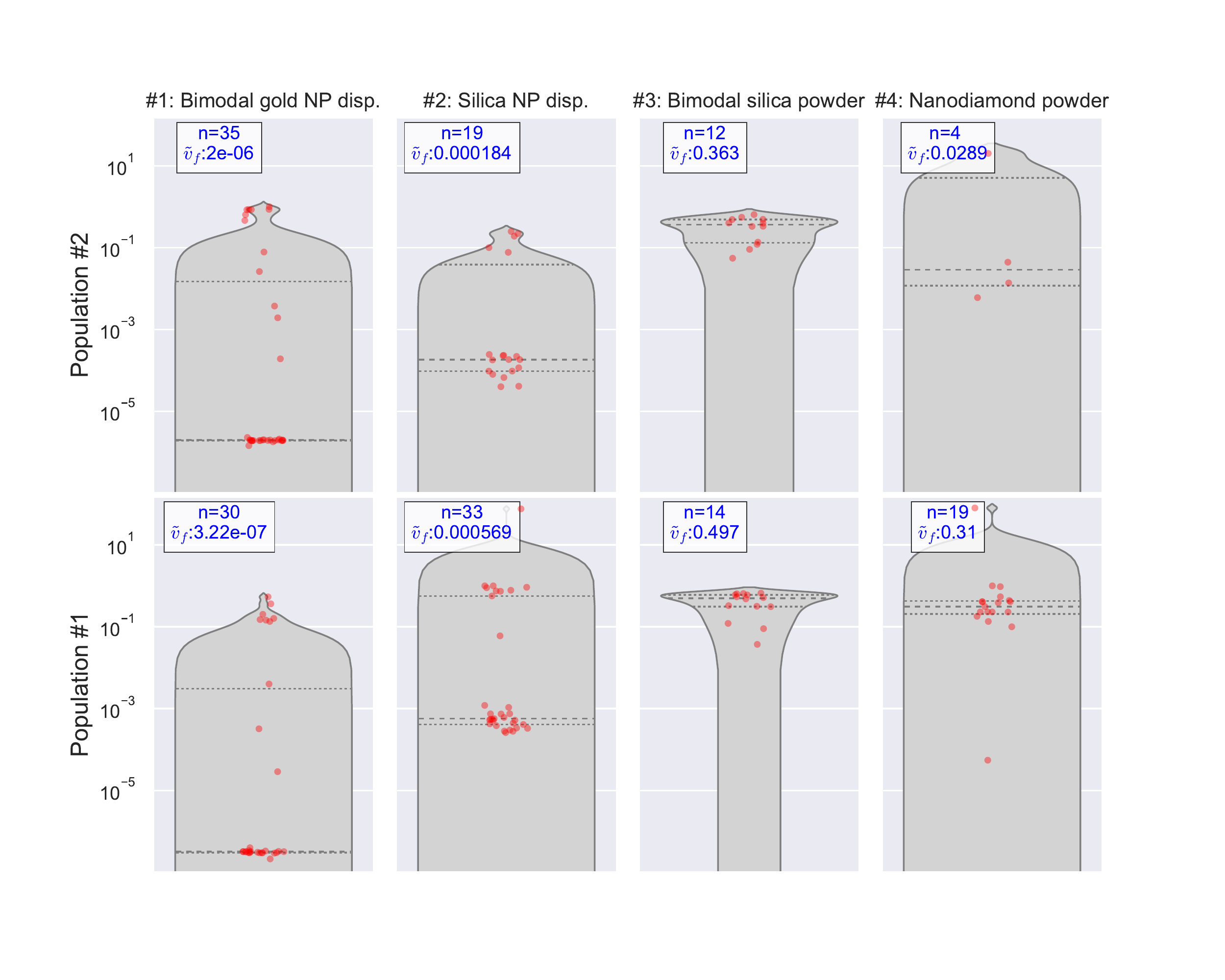}
\caption{The volume fractions reported by the participants, for each population of each dataset. The violin plot shows the quartiles as dashed lines, the red dots show the entries. The median volume fraction is shown in addition to the number of reported volume fractions in the text box in each plot.}
\label{fg:volfracs}
\end{center}
\end{figure}

Participants were asked to enter information on the volume fractions for each population where available. To enable this determination, the data was scaled to absolute units \protect{\footnote{Though, perhaps, using an incorrect thickness for the diamond powder and bimodal silica samples, as the apparent thickness of the materials were used based on their X-ray absorption, rather than the thickness of the actual container in which they resided.}}. 

While volume fractions are unambiguously defined, the results do not reflect this, showing an unusably wide spread in submitted values particularly for the dispersions (Figure \ref{fg:volfracs}). The origin of the spread, and thus the path through which they can be corrected, is not immediately clear. The sole but unsatisfactory conclusion is that there is unifying work to be done as well as cross-checks on how the volume fractions are computed and presented.

\section{A word on self-assessed experience}

\begin{figure}[ht]
\begin{center}
\includegraphics[width=.50\textwidth]{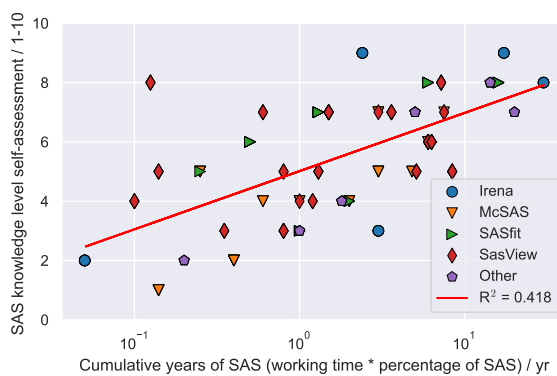}
\caption{The degree of SAS knowledge as a function of the cumulative SAS experience, showing a weak correlation as expected.}
\label{fg:betterwithyears}
\end{center}
\end{figure}

Multiple information fields were provided which at least tenuously link to the experience of the participant. These are the working years, the percentage of SAS in their working life, and a self-assessment of their level of knowledge (on a scale of 1 to 10). While this information offers only a very crude quantification of the scattering career of each individual, we can attempt to derive some insights from this \protect{\footnote{Missing information, for example, includes the field of expertise of the participants, the changes in SAS fractions of their daily routine over the years, experience with analysis in particular, etc.}}. As is to be expected (Figure \ref{fg:betterwithyears}), the self-assessed level of knowledge does correlate weakly with cumulative years of working with SAS, calculated as the product of the percentage of SAS in their working life with the years of their working life. In other words, the longer and more one works in the field, the higher their estimate of their working knowledge. 

\begin{figure}[ht]
\begin{center}
\includegraphics[width=.99\textwidth]{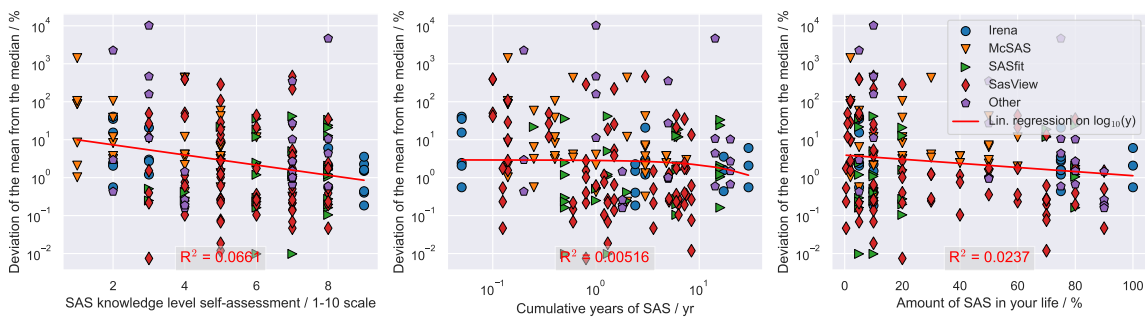}
\caption{Weak evidence of correlations, between several experience measures (SAS knowledge, cumulative years of SAS, and percentage of SAS in life) and the closeness to the median result. The results unfortunately show little to no correlation. }
\label{fg:better}
\end{center}
\end{figure}

It is perhaps to be expected that some of these measures would correlate strongly with the closeness of the result to the median results, assuming that the median is an proximate to the target values. Figure \ref{fg:better} shows a that such a correlation is, if present, only weakly present. This is unfortunate, as it would imply that we are not automatically getting better with more experience. One explanation could be that the true genius of the participants is being held back by both the limitations in the reporting by the software, as well as the mentally taxing needless dichotomies found in the field (c.f. Paragraph \ref{sc:suggestions}). It seems, then, that in order to improve as a community, we need to do more than merely getting older. 


\section{Potential steps for immediate improvement}\label{sc:suggestions}

Apart from the population means for the dispersions, the large spread of the remaining population parameters found in this work highlights that the human factor has the potential to introduce a significant uncertainty into the overall small-angle scattering data interpretation process. This uncertainty, as estimated in this study, is much larger in magnitude than those arising from data collection and corrections alone \cite{Pauw-2017a, Smales-2021, Schavkan-2019}. Some of the difficulties associated with the interpretation of scattering data start when researchers are faced with a barrage of possible units, non-standardized data formats, and poorly specified data practices even before analysis can begin. Expecting unfamiliar users to gain an in-depth understanding of the various redundant units in circulation, in addition to the pits and falls of each analysis method, \emph{and} understanding the differences in their implementations in respective fitting software forms a high barrier of entry. A further problem is that this barrier of entry appears invisible to many within the community (or worse: is considered a rite of passage), all of which can easily lead to unsatisfactory interpretations.

To alleviate this, our community should refrain from actively confusing users through a lack of constraint and definition. In other words, instrument responsibles in collaboration with software developers have to agree on -- and themselves adhere to -- a consistent set of units and definitions. Secondly, universal guides should be established (perhaps by a CanSAS or IUCr working group) on how to approach data analysis challenges of common sample types, rather than relying on local knowledge transfer alone. Lastly, users of software packages should take some time to read software documentation and understand the values the software is presenting. Conversely, software documentation can be written to contain easy-to-understand sections \cite{Wuttke-2022}.

Thus, immediate improvements in inter-laboratory result consistency may be obtained through:
\begin{enumerate}
\item{} Gradually aligning the information reported by the various software packages, e.g., presenting universal population information in the form of distribution moments (total value, mean, variance, skew and kurtosis) for each population
\item{} Providing user guides for approaching standard scattering analysis problems, providing robust model suggestions and adaptation approaches for dilute as well as dense systems. Additional methods for rough estimate cross-checks and result validation should be provided as well, i.e., make sure the dimensions are commensurate with the q range etc.. 
\item{} Introducing and using practically reasonable data uncertainty estimates (e.g., using methods used in \cite{Smales-2021}) in fits, so that reliable datapoints weigh more heavily than unreliable datapoints, which incidentally will also result in 
\item{} The provision of consistent and comparable goodness-of-fit measures to qualify a fit. With good uncertainty estimates, these goodness-of-fit measures will also have meaning.
\item{} Removing trite, time-consuming yet unnecessarily confusing dichotomies and the risks of errors in the therefore required unit conversions \protect{\footnote{Although the mechanism by which any community may agree on one of two options is itself a veritable wasp's-nest of conflict}}. A non-exhaustive list could be:

\begin{itemize}
    \item default units of Q should be defined (e.g. $nm^{-1}$)
    \item default units of I should be defined (e.g. $(m \cdot sr)^{-1}$)
    \item size should consistently refer to the full-length (diameter) of objects instead of occasionally referring to the half-length (radius) for select shapes. This way, mistakes in factor-of-two shifts are avoided when moving to other scatterer shapes.
    \item population information should be either volume-weighted for closer reflection of the information content of a scattering pattern, or number-weighted, but whichever it is, it has to be \emph{clearly} and repeatedly indicated.
\end{itemize}
\end{enumerate}




\section{Tips for future Round Robin experiment designs}

No experiment is perfect, and this Round Robin is no exception. Future iterations may include the following improvement suggestions. 

One way to bring together a larger community and gain insight into the progression (or regression) of the agreement, is to stage regular, smaller Round Robin studies. This could be as straightforward as providing one dataset per semester. This has the added advantage of building up a library of data and fit examples.

Further separation of the effects of the user vs. that of the software, will help to identify the main source of uncertainty. To that end, some Round Robin studies could dictate the use of a particular software package, or a particular model. 

Cross-evaluation of the quality of fits might allow an assessment on the level of agreement on what constitutes a ``good fit''. This can also lead to the identification of the target or best fit to compare against. Following on this, all necessary metadata required to reproduce a fit should be preserved by each participant, so that sources of disagreement can be better identified once a consensus fit has been established. 

\section{Conclusions}

A Round Robin study has been carried out to study the effect of individual researchers on the numerical results of a small-angle scattering pattern analysis. Before analysis, several results required corrections to compensate for field-specific dichotomies resulting from omitted or incorrectly applied unit conversions in ingestion as well as reporting. 

The results highlight a narrow spread in determined population means for samples consisting of low-concentration dispersions of globular scatterers, with half of the entries falling within 1.5\,\% of the median mean. For more challenging scattering patterns of concentrated powders, the spread is considerable to excessive, with half of the entries within 44\,\% of the median mean. This is likely due to the results being additionally affected by the choice of structure factor model and volume fraction. The determined population widths for both types vary wildly, and are ostensibly incomparable due to the differences in parameters that are reported by the various software packages (this, despite the nominal answer format specifying specifically a width in the form of a standard deviation). Lastly, considerable confusion exists on whether some software packages report fitting parameters as volume- or number-weighted values. 

Additionally, while participants do estimate their knowledge to be higher the longer they work with the method, this does not strongly correlate to a closer proximity to the median means. Therefore, alternative suggestions (i.e., besides acquiring years of professional experience) are provided in paragraph \ref{sc:suggestions} that could help improve the intercomparability of obtained results, in particular for widths and volume fractions. The implementation of a subset of these is bound to have a positive effect on the comparability of scientific results obtained with small-angle scattering.

\section{Data availability}

The data as well as the Jupyter notebook used for data sanitizing, analysis and graphing, is available on the Zenodo open-access data repository \cite{DARR-2023,DARRdataset-2023}. Further analysis and extension of this data is strongly encouraged.

\section{Acknowledgments}

K.T. acknowledges the NIST-NRC postdoctoral fellowship program for support. Certain commercial equipment, instruments, materials, or software are identified in this paper in order to specify the experimental procedure adequately. Such identification is not intended to imply recommendation or endorsement by NIST, nor is it intended to imply that the materials or equipment identified are necessarily the best available for the purpose. This work was partially funded through the European Metrology Programme for Innovation and Research (EMPIR) project No. 17NRM04.

\section{Bibliography}

\bibliographystyle{vancouver}
\bibliography{main}














\end{document}